# Extension of the bright high-harmonic photon energy range via nonadiabatic critical phase matching

**Authors**

Zongyuan Fu,[1][†] Yudong Chen,[1][†] Sainan Peng,[1] Bingbing Zhu,[1] Baochang Li,[2] Rodrigo Martín-Hernández[3], Guangyu Fan,[4] Yihua Wang,[1] Carlos Hernández-García,[3] Cheng Jin,[2][*] Margaret Murnane,[5] Henry Kapteyn,[5] Zhensheng Tao[1][*]

**Affiliations**

[1]Department of Physics and State Key Laboratory of Surface Physics, Fudan University, Shanghai 200433, China.
[2]Department of Applied Physics, Nanjing University of Science and Technology, Nanjing, Jiangsu 210094, China.
[3]Grupo de Investigación en Aplicaciones del Láser y Fotónica, Departamento de Física Aplicada, Universidad de Salamanca, E- 37008, Salamanca, Spain
[4]Institut National de la Recherche Scientifique, Centre Énergie Matériaux et Télécommunications, Varennes, Quebec, Canada.
[5]Department of Physics and JILA, University of Colorado and NIST, Boulder, CO 80309, USA.
[*]email: cjin@njust.edu.cn
[*]email: zhenshengtao@fudan.edu.cn
[†] These authors contributed equally to this work.

**Abstract**

Extending the photon energy range of bright high-harmonic generation to cover the entire soft X-ray region is important for many applications in science and technology. The concept of critical ionization fraction has been essential, because it dictates the maximum driving laser intensity that can be used while preserving bright harmonic emission. In this work, we reveal a second, nonadiabatic critical ionization fraction that substantially extends the maximum phase-matched high−harmonic photon energy, that arises due to strong reshaping of the intense driving laser field. We validate this understanding through a systematic comparison between experiment and theory, for a wide range of pulse durations and driving laser wavelengths. In particular, high harmonics driven by intense few-cycle pulses experience the most pronounced spectral reshaping, significantly extending the bright photon energy range. We also present an analytical model that predicts the spectral extension that can be achieved for different driving lasers. This reveals an increasing role of nonadiabatic critical phase matching when driven by few-cycle mid-infrared lasers. These findings are important for the development of high-brightness soft X-ray sources for applications in spectroscopy and imaging.

# MAIN TEXT

## Introduction

High-harmonic generation (HHG) in atoms and molecules(*1–4*) has enabled many advances in attosecond science and technology(*5–7*). HHG arises from the interaction of an intense laser field with an atomic or molecular gas target. The underlying process at the atomic and quantum level is based on strong-field ionization of an atom - a laser field modulates the spatial and temporal wavefunction of the outermost electron, leading to a rapidly changing dipole moment and emission of high harmonics of the driving laser field(*8, 9*) . Simple predictions of the maximum photon energy can be made using a semiclassical model(*10, 11*). In this picture, the liberated electron is accelerated by a laser field, and can be driven back to recollide with the parent ion when the laser field reverses, generating an extreme ultraviolet (XUV) or soft X-ray photon. The maximum photon energy achievable is determined by the maximum kinetic energy of the returning electron, which is given by the cutoff energy(*11*): $E_{cutoff} = I_p + 3.17U_p$. Here, $I_p$ is the ionization potential and $U_p$ is the ponderomotive energy, $U_p(\text{eV}) = 0.0933 I_L(\text{TW cm}^{-2})\lambda_L^2(\mu m^2)$, where $I_L$ and $\lambda_L$ are the peak intensity and wavelength of the driving field.

Extending the bright photon energy range of HHG is important for many applications in science and technology: first, bright and broadband soft X-ray radiation spans many element-specific absorption edges, e.g. the "water-window" energy(*12*), and enables table-top element-resolved spectroscopy and microscopy(*13–16*). And second, broader HHG bandwidths can support the generation of shorter attosecond pulses (*17, 18*), or powerful broadband imaging and spectroscopy methods(*19*). However, the HHG emission from a single atom exhibits a dipole pattern and is very weak. In order to generate a bright and directed high harmonic beam, the driving laser and the generated harmonics must both travel at the same phase velocity in the gas, so that the HHG emission from many atoms combines constructively. Here, the concept of critical ionization fraction ($\eta_c^{PMC}$) is essential (*1, 20–22*). Only when the laser-induced gas ionization is lower than a critical ionization level $\eta_c^{PMC}$, the phase-mismatch between the fundamental driving field and the harmonic field can be compensated, allowing for coherent build-up of HHG radiation along the propagation direction. Typical critical ionizations levels $\eta_c^{PMC}$ in a gas medium range from <5%, which places a limit on the maximum driving laser intensity for which phase-matching is possible, and thereby determines the phase-matching cutoff (PMC). Although more exotic schemes can support effective phase matching at very high levels of ionization(*23*) , simultaneously achieving high photon energy and high brightness is still challenging.

Driving HHG with intense few-cycle pulses has attracted interest, because it is one route for generating isolated attosecond pulses(*17, 18, 24–28*). When driven by few-cycle pulses, the maximum HHG photon energy was reported to be greatly extended(*29–31*). This is not only because pulse compression can boost the laser intensity, but also because it facilitates phase matching for high-frequency harmonics by reducing gas ionization – for example, by reducing the pulse duration from 20 fs to 5fs, the phase matching cutoff can be extended by ~20%. However, to understanding the HHG spectrum beyond the PMC also needs to consider the significant reshaping of an intense few-cycle pulse in a dispersive plasma (*30–33*). In Fig. 1A, we plot a simple and intuitive schematic of HHG in such regime, which involves several effects, such as an intensity drop due to plasma-induced defocusing, deformation of the laser field, a longitudinal gradient of gas ionization, etc. The emitted harmonic spectrum is the result of the coherent addition of the emission from numerous emitters along the propagation direction. Thus, an important question to address is what is the maximum achievable HHG photon energy under such nonadiabatic conditions, that supports a practically usable source brightness i.e. reasonable phase matching?

In recent years, there has been renewed interest in studying nonadiabatic effects in HHG, owing to advances in high-energy few-cycle mid-infrared lasers(*14*, *34*). Experimentally, it has been shown that the plasma- and nonlinear pulse reshaping is important for both several-cycle 4 µm lasers, or few-cycle 1.8-µm lasers, for the generation of high-flux >1000 eV and 600-eV soft X-ray harmonics respectively(*34*, *35*). Experimentally, the sub-cycle deformation of the driving field occurs in many perturbative and non-perturbative nonlinear-optical processes, and can be directly measured with the attosecond streaking technique(*36*). Theoretically, several important aspects of nonadiabatic effects have been revealed. It was shown that the plasma defocusing can clamp the laser intensity, limiting the harmonic energy(*37*). It has also been reported that the electronic trajectories from non-adiabatic drivers are modified (*38*). In addition, plasma-induced reshaping of the driving field can modulate sub-cycle electron trajectories, supporting phase-matching of high-energy harmonics (*39*). A similar mechanism has been used to explain experiments with harmonic orders significantly higher than PMC(*33*, *40*). Generation of isolated attosecond pulses is also possible(*24*, *41*, *42*). It is worth noting that, thus far, there is no simple and analytical model that can guide the development of a phase-matched harmonic source under nonadiabatic conditions, in contrast to adiabatic HHG, where several successful models been developed (see for example(*1*, *20*, *43*, *44*)).

In this work, we introduce the new concept of nonadiabatic critical ionization fraction (NCIF) that explains the reshaping and extension of a phase-matched harmonic spectrum under nonadiabatic conditions. Our experimental and theoretical results demonstrate that the concept of NCIF is crucial to understand and achieve nonadiabatic phase-matched HHG effects, such as the extension of the spectral roll-offs beyond the PMC. Special attention has been paid to HHG driven by intense few-cycle pulses, which can produce the highest roll-off-energy extension. In particular, when driven by 9-fs, 1030-nm and ~400 TW cm$^{-2}$ peak intensity laser pulses in argon, the harmonic spectrum extends to ~125 eV, which is ~50 eV higher than the corresponding PMC. Moreover, the brightness is between $10^9$ and $10^7$ photons/s/eV, which is sufficient for many applications in spectroscopy, and compares well with a peak brightness of >$10^{10}$ photons/s/eV at lower photon energies under perfect phase matching conditions. We identify three intensity regimes separated by two critical driving intensities, which determines the harmonic spectral shape. We have developed an analytical NCIF model that can precisely predict the spectral reshaping and extension under different driving-laser intensities, wavelengths, and pulse durations, and in different gas species. Furthermore, our model also reveals that the spectral extension can be much greater when driven by long-wavelength, mid-infrared few-cycle lasers. Our results provide an alternative route guideline for extending the harmonic frequency to >1 keV, well into the soft X-rays, with high brightness placing nonadiabatic HHG as a practical soft X-ray source for applications in ultrafast spectroscopy and imaging (*34*, *45*, *46*).

**Results**

**Harmonic spectrum roll-offs under different driving conditions**

The picture of macroscopic nonadiabatic HHG is depicted in Fig. 1A. When the driving laser peak intensity $I_L$ is high, beyond the tunnel ionization regime, nonadiabatic effects appear in HHG. In the intensity regime we are interested, where the target atoms are not fully ionized, the driving field is reshaped when propagating through the gas cell as follows (*47*). First, the rear part of the pulse undergoes spatial defocusing due to gas ionization induced during the earlier cycles, which leads to a peak intensity drop and a shift of the envelop peak to the leading edge (see the inset of Fig. 1A). Second, the rapid sub-cycle variation of gas ionization causes a strong frequency blueshift of the pulse leading edge (*43*) (see the inset of Fig. 1A).

In order to investigate how such propagation of the fundamental field affects phase-matching in HHG, we conducted systematic experiments under different laser conditions and in different gas species. In our experiments, laser pulses with full-width-at-half-maximum (FWHM) temporal durations of $\tau$=9, 22 and 170 fs, and peak intensities above 180 TWcm$^{-2}$, at a center wavelength of $\lambda_L$=1030 nm are focused into a gas cell filled with argon and krypton. The gas-cell length is $d$=1.5 mm. The high harmonic spectrum is recorded by an XUV spectrometer after filtering out the fundamental driving laser with metal thin films. The details of the experimental setup are summarized in Materials and Methods and in Supplementary Materials (SM). In Fig. 1C-E, we plot the experimental harmonic spectra in argon driven by $\tau$=9, 22 and 170 fs pulses, respectively, and for different intensities. Here, we note that the reported laser intensities ($I_L$) in this work are those at the entrance of the gas cell, which can be reduced by plasma-induced defocusing inside the gas cell (see Fig. 1A). The gas pressures $p$ are 50, 190 and 210 torr, respectively for the three cases.

Under such high-intensity conditions, as the harmonic spectrum is obtained by summing up all the emitters along the propagation direction ($z$) under a varying field intensity (as illustrated in Fig. 1a), the classic microscopic cutoff energy ($E_{cutoff}$) is no longer discernible. To quantitatively analyze the harmonic spectral shape, we focus on two special harmonic energies (as labeled in Figs. 1C-D): i) the phase-matching cutoff energy ($E_{PMC}$), which corresponds to the energy from which the harmonic yield continuously decreases (*35*); and ii), the 1%-intensity energy ($E_{1\%}$) where the spectral intensity decreases to 1% of the intensity at $E_{PMC}$. The energy difference $\Delta E = E_{1\%} - E_{PMC}$ represents the width of the spectral roll-off beyond the $E_{PMC}$. As shown in Fig. 1C ($\tau$=9 fs and $I_L$=360 TW cm$^{-2}$), $\Delta E$ can be as large as ~50 eV, which is comparable to the corresponding $E_{PMC}$ (~75 eV), delivering a great amount of usable high-energy XUV photons beyond the PMC. The flux at ~125 eV can reach ~2×10$^7$ photons/s/eV. In contrast, the flux at the same energy is reduced by more than two orders of magnitude when driven by 22-fs, 400 TW cm$^{-2}$ pulses, becoming not measurable with the 170-fs, 180 TW cm$^{-2}$ pulses. Indeed, for the $\tau$=22 fs and 170 fs cases, $\Delta E$ is reduced to ~21 eV and ~14 eV, respectively.

In order to investigate the physics beyond the experimental results, we have performed numerical simulations under similar conditions, as shown in Figs. 1F-H. The simulations were performed within the strong-field approximation (SFA) framework, taking into account the high-intensity regime (see Materials and Methods). The ionization rates have been calculated through an empirically modified Ammosov-Delone-Krainov (ADK) model(*48*) to take into account laser intensities above the tunnel ionization regime. Furthermore, the agreement between the SFA and the time-dependent-Schrödinger-equation (TDSE) in the single-atom-simulation results further indicates that barrier suppression effects have limited influence in the intensity range of our experiments (see Supplementary Fig. S7). Interestingly, we find that, when the driving intensity $I_L$ is low (e.g. $I_L$ <200 TW cm$^{-2}$ for $\tau$=9 fs), $\Delta E$ is typically 10~14 eV, which is independent to the laser conditions (duration, intensity, wavelength, etc.) and the gas species (see SM). Furthermore, we find that this value is consistent with the single-atom response (see Supplementary Fig. S6 and S7), indicating that it represents the quantum limit on the minimum energy extension ($\Delta E_{qt}$). Thus, the deviation of $\Delta E$ from $\Delta E_{qt}$ can be attributed to the influence of macroscopic phase-matching and nonadiabatic effects.

Since the reshaping of the harmonic spectrum is most pronounced when the driving pulse duration is short, we now focus our attention on the results of $\tau$=9 fs. In Fig. 2 we show the influence of the gas plasma on the driving field, by plotting the simulation results of the driving field pulse at the entrance (black) and at the exit (red) of the gas cell for low (A, 200 TW cm$^{-2}$) and high (B, 700 TW cm$^{-2}$) values of $I_L$. From the comparison of Fig. 2A and B, we find that the driving field is strongly reshaped by the gas plasma in the high-intensity regime, resulting in a decrease in the field intensity, a strong frequency blueshift at the pulse leading edge, and a shift of the envelop peak to the leading edge by about one optical cycle. The time-frequency analysis on the numerical results of the corresponding HHG spectra are further shown in Fig. 2C and D. Correspondingly, the high-energy harmonic photons are dominantly generated at about one optical cycle before the pulse temporal center when the driving laser intensity is high (Fig. 2D). This result clearly demonstrates that the nonadiabatic effects play an essential role in the generation and phase matching of the high-energy harmonics beyond PMC. We note that the propagation effects have been considered in the numerical simulations of the HHG spectra. The spectral results are obtained by integrating across the transverse beam cross-section.

To what extent the harmonic energy can be extended by the high laser intensities? To address this question, we measure the HHG spectrum driven by increasing $I_L$ with $\tau$=9 fs in argon (Fig. 2E) and krypton (see SM) with $d$=1.5 mm and $p$=50 torr. Here, we will first restrict our discussion to the case where $d$ and $p$ are optimized for the bright HHG emission around $E_{PMC}$. The situations with varying $d$ and $p$ will be discussed later. The energies of $E_{PMC}$ and $E_{1\%}$ can then be labeled in the same way as illustrated in Fig. 1C-E. Interestingly, as shown in Fig. 2E, we can clearly distinguish three intensity regimes separated by two critical intensities ($I_c^{PMC}$ and $I_c^{NCIF}$): (i) When $I_L < I_c^{PMC}$ and the gas ionization is low, the deformation of the driving field is negligible (Fig. 2A), and thereby, $E_{PMC}$ is in excellent agreement with $E_{cutoff}$ (the blue-dashed line). Furthermore, $\Delta E$ in this regime equals $\Delta E_{qt}$, and, as a result, $E_{1\%}$ also increases linearly (the yellow dash-dot line), simply following $E_{1\%} = \Delta E_{qt} + E_{cutoff}(I_L)$. (ii) When $I_L > I_c^{PMC}$, the gas ionization is high enough to deform the driving field upon propagation. We identify $I_c^{PMC}$ as the threshold intensity from which nonadiabatic effects play an important role. In this regime, we find that $E_{PMC}$ saturates at the energy of $E_{PMC}^{sat}$ ~75 eV, which corresponds to $E_{cutoff}$ at $I_L = I_c^{PMC}$. In contrast to $E_{PMC}$, $E_{1\%}$ continues to increase almost linearly in this regime, until $I_L$ reaches the next critical intensity $I_c^{NCIF}$ (~300 TW cm$^{-2}$). Remarkably, in this regime, the harmonic roll-off, given by $\Delta E$, continue to increase with $I_L$, leading to the most pronounced spectral reshaping. But, (iii) when $I_L > I_c^{NCIF}$, our results show that $E_{1\%}$ eventually saturates at the energy of $E_{1\%}^{sat}$ (~125 eV for $\tau$=9 fs in argon), which corresponds to the sum of $\Delta E_{qt}$ and $E_{cutoff}$ when $I_L = I_c^{NCIF}$. We note that the appearance of regime (iii) also indicates the saturation of the spectral reshaping, and further increasing the harmonic energy becomes not possible. As shown in Fig. 2F, these observations can be quantitatively reproduced by the numerical simulations, including the values of $I_c^{PMC}$ and $I_c^{NCIF}$. Moreover, the above three regimes can be universally observed under different pulse durations, gas atoms and wavelengths (see SM). But, the critical intensities and the corresponding saturation energies can be altered by these conditions.

**The NCIF model**

As illustrated in Fig. 1A, in the high-intensity regime we are analyzing, the high-energy HHG photons can only be generated at the early stage of the driving-laser propagation, before the laser intensity is reduced by plasma defocusing. In this region, the sub-cycle deformation of the driving field can play an essential role in phase matching (*36, 39, 49*). The wavevector mismatch between the fundamental driving field and the $q^{th}$-order harmonic field can be expressed as a sum of four terms (*1, 43*):

$$\Delta k_q = \Delta k_g + \Delta k_n + \Delta k_p + \Delta k_d. \tag{1}$$

Here, $\Delta k_g$ denotes the geometrical Gouy-phase term due to focusing. $\Delta k_n$ and $\Delta k_p$ account for the pressure-dependent contributions from the neutral-atom and free-electron dispersions, respectively, and are given by:

$$\Delta k_n = \frac{2\pi q}{\lambda_L} \frac{p}{p_{atm}} [1-\eta(t)]\Delta\delta, \tag{2}$$

$$\Delta k_p = -q \frac{p}{p_{atm}} r_e \eta(t) N_{atm} \lambda_L (1-\frac{1}{q^2}), \tag{3}$$

where $p_{atm}$ is the atmospheric pressure, $\eta(t)$ is the ionization fraction, $\Delta\delta$ is the difference between the indices of refraction of the neutral gas per atmosphere at the fundamental and X-ray wavelengths, $r_e$ is the classical electron radius, and $N_{atm}$ is the number density per atmosphere. Finally, $\Delta k_d$ represents the wavevector mismatch induced by the dipole phase (*50*). Under the adiabatic conditions, this term is usually small for the phase-matching of short-trajectory harmonics (*1*, *51*). Under the nonadiabatic conditions, however, due to the strong reshaping of the laser field under propagation, this term can become especially relevant for the phase-matching of high-energy harmonics(*36*, *39*, *49*). Here, we derive that $\Delta k_d$ is dominantly contributed by the phase mismatch due to the frequency blueshift of the driving field, which is the result of the sub-cycle variation of gas ionization ($\frac{\partial \eta}{\partial t}$)(*52*) by

$$\Delta k_d \approx \alpha_j \frac{3U_p \lambda_L^3}{2\pi hc^2} \frac{p}{p_{atm}} r_e N_{atm} \frac{\partial \eta(I_L,t)}{\partial t}, \tag{4}$$

where $\alpha_j$= 3.2 represents the phase coefficient at the cutoff energy (*43*), $h$ is Planck constant and $c$ the speed of light. The details of the derivation of Eq. (4) are provided in Supplementary Section S4.

Since $\Delta k_g$ is negative and independent to the pressure, phase matching of HHG is only possible when $\Delta k_n + \Delta k_p + \Delta k_d > 0$. Hence, a critical ionization fraction including the nonadiabatic effects can be derived:

$$\eta_c(I_L) \approx \frac{\frac{2\pi}{\lambda_L}\Delta\delta + C_d \frac{\alpha_j}{2\pi c} \frac{3U_p}{I_p + 3.17U_p} \lambda_L^2 r_e N_{atm} \frac{\partial \eta(I_L,t)}{\partial t}\big|_{t=0}}{\frac{2\pi}{\lambda_L}\Delta\delta + r_e N_{atm} \lambda_L}. \tag{5}$$

Here, we focus on the cutoff energy with $E_q = q\frac{hc}{\lambda_L} = I_p + 3.17U_p$, and introduce a parameter $C_d$, with which we can control the contribution of $\Delta k_d$. Under the critical condition, we have another constraint that the ionization fraction at the peak of the pulse should just reach $\eta_c$:

$$\eta_c(I_L) = \int_{-\infty}^{0} \frac{\partial \eta(I_L,t)}{\partial t} dt, \tag{6}$$

and Eq. (6) can be calculated by the ADK theory (*53*) (see SM).

Taking argon as an example, as shown in Fig. 3A, Eqs. (5-6) are solved by finding the points of intersection of the two curves for different pulse durations, which allows to extract the critical intensities ($I_c^{PMC}$ and $I_c^{NCIF}$) and the corresponding critical ionization fractions ($\eta_c^{PMC}$ and $\eta_c^{NCIF}$). Obviously, when $C_d = 0$ (solid black line), the solution of Eqs. (5-6) is simply reduced to $\eta_c^{PMC}$, and the intersections (open triangles) of $\eta_c^{PMC} \approx 2.2\ \%$ are in excellent agreement with the previous studies(35, 54). The predicted $I_c^{PMC}$ and $E_{PMC}^{sat}$ are also in agreement with the experimental results as shown in Fig. 3B (also see SM). The solutions for $\eta_c^{NCIF}$ and $I_c^{NCIF}$ under different driving pulse durations are further given by the intersections of the two curves with $C_d \neq 0$ (open squares). Here, an energy upshift of $\Delta E_{qt} \approx 12$ eV is necessary for the agreement of $E_{1\%}^{sat}$, as illustrated in Fig. 3A. With the above method, we can apply the NCIF model to fit the experimental results with only one free parameter, $C_d$. As shown in Fig. 3B and C, remarkably, we find that with $C_d = 0.35$ the model can well reproduce the experimental results under different pulse durations and in different gas atoms (argon and krypton) (see SM).

The NCIF model also explains why there is a limit to the spectral extension under high driving laser intensity, as shown in Fig. 2E and F. Under the nonadiabatic conditions, the transient phase matching process can contribute a large amount of positive wavevector mismatch to compensate the negative contribution from the plasma dispersion in Eq. (1), supporting the generation of harmonic orders well above the PMC when the driving laser field is strong and the laser-induced ionization level is high. However, this phase-matching capability has its limit. When $I_L > I_c^{NCIF}$, the gas ionization becomes so high that this compensation is no longer possible. This critical point corresponds to $\eta_c^{NCIF} \approx 13\%$ for $\tau$=9 fs and $\lambda_L$=1030 nm, which is several times higher than $\eta_c^{PMC}$ in argon. This explains the large energy extension of $\Delta E$~50 eV (Fig. 1C). When driven by the longer pulses, $\eta_c^{NCIF}$ are greatly reduced ($\eta_c^{NCIF} \approx$3.8% for $\tau$=22 and $\approx$2.4% for $\tau$=170 fs), as shown in Fig. 3A. The reason is twofold: first, the accumulated gas ionization rises faster for the longer pulses, and second, the time derivative of the gas ionization ($\frac{\partial \eta}{\partial t}$) also reduces when the ionization is high (see SM), which makes the bending down of the curve for Eq. (6) to occur at lower intensities (Fig. 3A).

The influence of the carrier-envelop phase (CEP) $\Delta\varphi_{CEP}$ is investigated with the NCIF model as shown in Fig. 3B and C. We find that the harmonic spectrum varies with a period of $\pi$ as a function of $\Delta\varphi_{CEP}$, consistent with the previous studies (34, 55, 56). The model shows that $\Delta\varphi_{CEP}$ has great effects when driven by few-cycle pulses, while its influence becomes negligible when the duration is longer than 5 optical cycles ($\tau$~17 fs for $\lambda_L$=1030 nm). Meanwhile, the energy of $E_{1\%}$ is much more strongly affected than $E_{PMC}$. Our results show that, when $\Delta\varphi_{CEP} = \frac{\pi}{2}$ with the peak field shifted to the pulse leading edge (inset of Fig. 3(B)), the ionization rate can be reduced, and thereby, the curves of Eqs. (5-6) cross at stronger laser intensity, generating higher-energy harmonics. In contrast, when $\Delta\varphi_{CEP}$ approaches $-\frac{\pi}{2}$, the ionization increases due to more optical cycles before the peak field, which yields lower critical intensity in the model. We note that, using the NCIF model, we reveal the phase-matching aspects of the CEP effects, but this model cannot explain the details of the spectral shapes under the modulation of electron trajectories (55). It also worth noting that the model results represent the upper limit of the $E_{PMC}$ and $E_{1\%}$ variations, when the laser intensity ($I_L$) is sufficiently higher than $I_c^{NCIF}$.

The fact that $C_d$<1.0 indicates that the nonadiabatic dipole-phase mismatch ($\Delta k_d$) is not fully taken into account. First of all, the choice of $C_d$=0.35 is related to the fact that we focus on the phase-matching condition of the 1%-intensity energy ($E_{1\%}$). For the 10%-intensity energy, the same fitting procedure yields $C_d \approx 0.27$ and lower $\eta_c^{NCIF}$ (see SM), which indicates more restrictive phase-matching conditions for lower harmonic orders. Secondly, $C_d$<1.0 may also be attribute to the spatial and temporal averaging effects which is not fully considered in our analytical model (*34*).

**Extension to mid-infrared few-cycle lasers**

Here, we extend the NCIF model to longer driving wavelengths and to the gas atoms of neon and helium, which are of great interest for generating high-energy soft X-ray harmonics (*34*, *45*, *46*, *55*, *57*, *58*). In Fig. 3D, we plot the predicted $E_{PMC}^{sat}$ and $E_{1\%}^{sat}$ by the NCIF model for different laser wavelengths ($\lambda_L$). The FWHM duration is fixed to be 2.6 cycles and $\Delta\varphi_{CEP} = 0$ in the calculation. To verify our model results, we conducted a series of numerical simulations with different $\lambda_L$ in neon and helium with $I_L > I_c^{NCIF}$ (solid symbols in Fig. 3D, and see SM). A typical HHG spectrum obtained from our numerical simulations is plotted in Fig. 3E. The gas pressure is set in the range of 100~500 torr for different gas atoms and driving conditions. The three intensity regimes as shown in Fig. 2(E) and (F) can also be ubiquitously observed under long-wavelength driving fields (see Supplementary Fig. S5). As shown in Fig. 3D, the NCIF model can yield excellent agreement with the numerical results. We find that the energy extension ($\Delta E$) increases monotonically as a function of $\lambda_L$. Remarkably, when driven by a $\lambda_L$~2 μm laser in helium, $\Delta E$ could reach ~1 keV, much higher than the corresponding $E_{PMC}$ (~570 eV), leading to an HHG spectrum exceeding 1.5 keV. This is because the contribution of the nonadiabatic dipole-phase term grows quadratically with respect to $\lambda_L$ [see Eq. (5)], which makes the nonadiabatic phase matching play an increasingly important role for longer driving wavelengths. Meanwhile, our model also predicts that, to generate and phase-match such high-energy harmonics, the driving laser intensity needs to exceed $I_c^{NCIF}$ ≈1300 TW cm$^{-2}$ (see Supplementary Fig. S12).

**Discussion**

In Supplementary Fig. S12, we summarize the results of $E_{1\%}$ observed in the previous experiments driven by few-cycle pulses with different $\lambda_L$ (*14*, *31*, *34*, *36*, *39*, *46*, *55*, *56*, *58*), and compare them with the NCIF model. Very interestingly, we find that, for $\lambda_L$ in the near-infrared region (800 and 1030 nm), the experimental $E_{1\%}$ agrees very well with the NCIF model, while, for the mid-infrared wavelengths ($\lambda_L$ >1500 nm), the results are in better agreement with $E_{PMC}^{sat}$ of the model. We believe this can be attributed to two reasons: First, it is still challenging to produce high-intensity few-cycle mid-infrared pulses to exceed $I_c^{NCIF}$ in the experiments. The typical field intensity reported in these works for $\lambda_L$>1500 nm is <500 TW cm$^2$, which is far below the predicted $I_c^{NCIF}$ (~1300 TW cm$^{-2}$ for helium and ~800 TW cm$^{-2}$ for neon, see Supplementary Fig. S12). Second, our model also shows that the phase-matching pressures for $E_{PMC}^{sat}$ and $E_{1\%}^{sat}$ can be increasingly different for the mid-infrared driving fields. For example, we estimate that the phase-matching pressure of helium at the $E_{PMC}^{sat}$ energy is ~2 atm when $\lambda_L$ is 800 nm. This pressure grows to tens of atm when the wavelength is increased to 4 μm, in agreement with the previous experiments (*35*). But, in contrast, because of the increasing contribution of the nonadiabatic dipole phase under long driving wavelengths, the phase-matching pressure for $E_{1\%}^{sat}$ decreases from 1 atm to 0.1 atm when $\lambda_L$ changes from 800 nm to 4 μm (see Supplementary Fig. S11). This makes it very challenging to simultaneously phase-match both energy regions under a constant gas pressure when the driving wavelength is long. Usually in the experiments, one may focus on the phase-matching of the $E_{PMC}^{sat}$ energy to yield the highest harmonic flux, and hence, the harmonic emission beyond PMC can be extinguished by the phase and group velocity mismatches (*59*) under the high gas pressure. In the future, special phase-matching techniques, including predesigned gas cells and gas-filled waveguides(*60*), which can provide inhomogeneous distribution of gas pressure, may be useful to solve this problem. Our results here deliver an optimistic message that it is possible to greatly enhance and extend the HHG brightness and energy to cover the entire soft X-ray range (100 eV – 5 keV) with further development of few-cycle, mid-infrared and high-intensity lasers. Recent great advances in power-scaling laser technology(*61–63*), few-cycle pulse-generation techniques(*30, 64–68*), as well as novel wavelength-scaling techniques(*69–72*), in combination could achieve this goal in the near future.

The deformation of HHG spectrum resulting from the driving-intensity variation along $z$ (see Fig. 1A) can be clearly demonstrated by the experiments with different medium lengths ($d$) and gas pressures ($p$), as shown in Fig. 4A. We find that, when $d$ is reduced from 1.5 mm to 0.7 mm at low pressure ($p$=25 torr), the propagation in the gas medium is not long enough that the laser intensity remains much higher than $I_c^{PMC}$ throughout the propagation (see Fig. 4C). The high ionization level precludes the phase matching for the harmonics around PMC~75 eV. The harmonic spectrum is reshaped and we can observe a subsequent upshift of $E_{PMC}$. We note that, although the pressure-length product is reduced only by a factor of ~8 here, the HHG yield at ~75 eV is reduced by approximately two orders of magnitude (Fig. 4A), highlighting the significant influence of the nonadiabatic effect on the HHG spectral shape. On the other hand, by increasing $p$ to 150 torr at $d$=0.7 mm, $E_{PMC}$ and the harmonic yield can be nearly recovered (Fig. 4A). In this case, our simulations show that the increased pressure accelerates the laser-intensity drop along $z$, making it reduced to around $I_c^{PMC}$ within the 0.7-mm propagation, and this facilitates the phase matching around $E_{PMC}$. More interestingly, we find that, under all these conditions, the brightness of HHG around $E_{1\%}$ is almost unaffected. This can be explained by the fact that, in all the three cases, $I_L$ can cross $I_c^{NCIF}$ within the gas medium lengths (see Fig. 4B and C), which produces these high-energy photons.

In summary, based on the systematic experimental and theoretical investigations of HHG under different driving conditions, we develop a NICF model taking the nonadiabatic effects on the HHG phase matching into account. The model can precisely predict the reshaping and extension of a harmonic spectrum, which is especially important when driven few-cycle mid-infrared lasers. Our results have potential for great impact and widespread use considering the recent great advances in high-energy few-cycle mid-infrared lasers as the driving sources of HHG.

**Materials and Methods**

**Experimental setup**

The schematic of the experimental setup is shown in the Supplementary Fig. S1 (see SM). Laser pulses with different pulse durations at a repetition rate of 10 kHz are obtained by compressing the fundamental 170-fs pulses with all-solid-state compressors, which utilize the soliton management in periodic layered Kerr media (*30, 73*). In all the experiments, the driving laser beam is focused to a waist ($w_0$) of 40-50 μm through a gas cell with an inner diameter of $d$=1.5 or 0.7 mm. The gas cell is typically ~1.5 mm behind the beam focus to ensure the phase matching of short trajectories(*74*). The harmonic spectrum is recorded by an XUV spectrometer after filtering out the fundamental driving laser with aluminum or zirconia thin films. The brightness of the harmonic orders is measured with a calibrated photodiode (see SM).

**Numerical simulation**

In our numerical simulations, the single-atom response is computed under the framework of the SFA theory. An empirically modified ADK model was implemented to take into account the ground-state depletion due to the barrier suppression effect (*48*). Furthermore, the quantitative rescattering (QSR) model was used to take into account the photo-recombination cross-

section(*75*). To calculate the propagation effect, the single-atom-induced dipole due to local electric field is then inserted into Maxwell's wave equations of high-harmonic field in which the dispersion and absorption of gas medium are included. Meanwhile, the fundamental driving laser pulse is propagated in the medium and its spatiotemporal waveform is reshaped because of the effects of diffraction, nonlinear self-focusing, ionization, and medium dispersion(*76*). Once the propagation equations of high-harmonic field and fundamental laser are numerically solved, one can obtain macroscopic HHG spectra by integrating high-harmonic yield over the exit plane of gas medium.

**Figures and Tables**

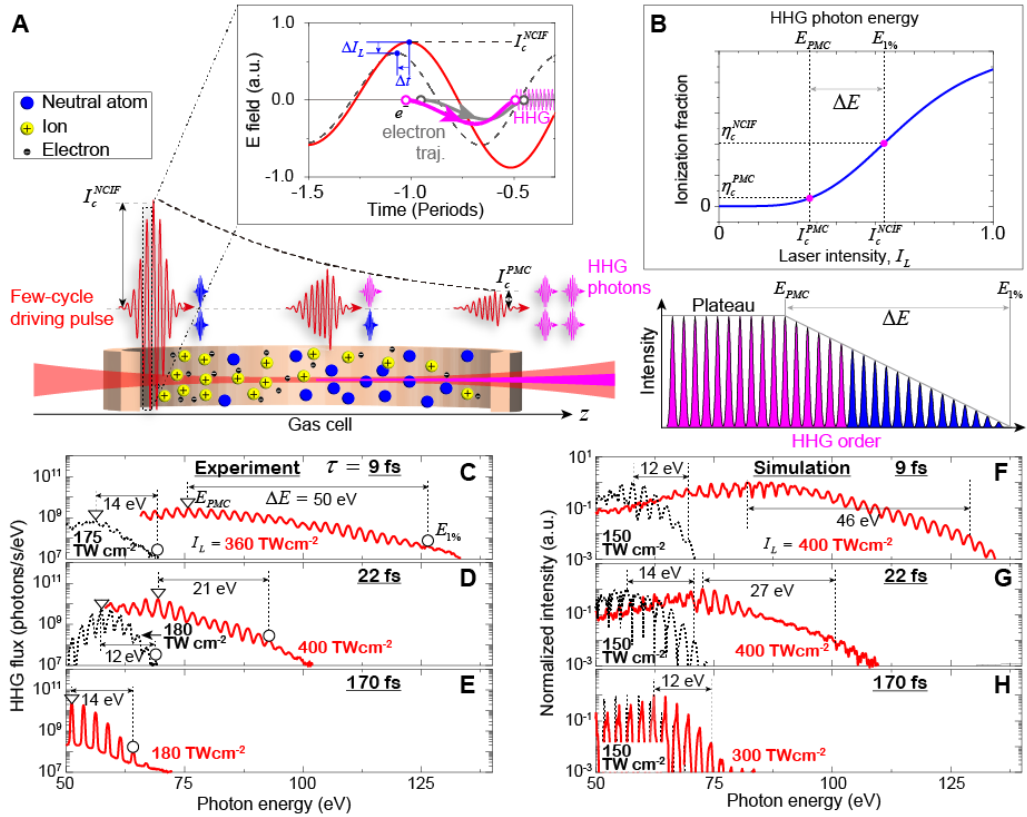

**Figure 1. Illustration of the general concept.** (A) Illustration of the nonadiabatic HHG process driven by an intense few-cycle pulse in a gas cell, which exhibits the effects of plasma-induced intensity decay, temporal pulse reshaping and variation of gas ionization along *z*. The HHG spectrum is the integrations of the emitters along *z* under a varying driving field. Inset: Illustration of the modulation of electron trajectories owing to the nonadiabatic field temporal reshaping. The solid line represents the incident driving laser field. The dashed line represents the field profile deformed by the nonadiabatic effects. The microscopic electron trajectories under two different driving fields are illustrated. (B) Illustration of the gas ionization fraction at the pulse temporal center, and the relationship between the critical intensities ($I_c^{PMC}$ and $I_c^{NCIF}$), critical ionization fractions ($\eta_c^{PMC}$ and $\eta_c^{NCIF}$) and the HHG photon energies ($E_{PMC}$ and $E_{1\%}$). (C)-(E) The HHG spectrum driven by pulse durations of $\tau$=9 fs, 22 fs and 170 fs, respectively, under different laser intensities ($I_L$). (F)-(H) The results of numerical simulations under similar conditions as in (C)-(E).

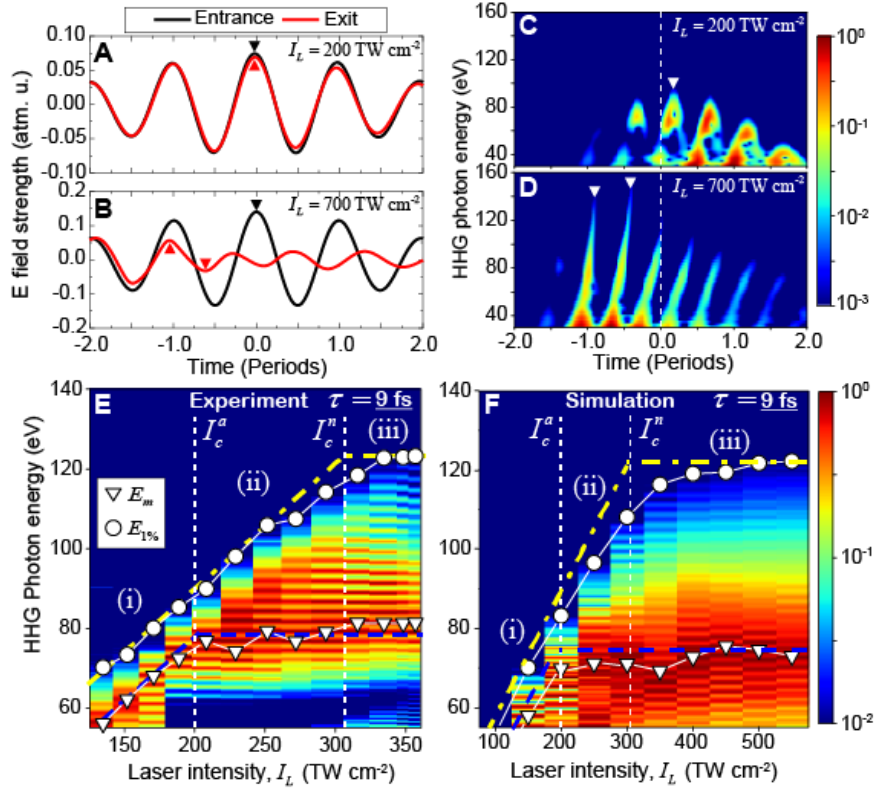

**Figure 2. Influence of nonadiabatic effects on harmonic spectrum.** (A) The temporal field shapes obtained from the numerical simulations at the entrance and the exit of a gas cell ($d$=1.5 mm) under laser intensity of $I_L$=200 TW cm$^{-2}$. The gas cell is filled with argon with the pressure ($p$) of 100 torr. The time for the peak field is labeled by the solid triangles. (B) Same as (A) for $I_L$=700 TW cm$^{-2}$. (C)-(D) The time-frequency analysis of HHG generated under the same conditions in (A) and (B). The emission time for the highest harmonic orders are labeled by the solid triangles. (E) The experimental HHG spectrum in argon driven by different laser intensity $I_L$ with $d$=1.5 mm and $p$= 50 torr. The pulse duration is $\tau$=9 fs. The blue dashed line represents $E_{cutoff}$ when $I_L < I_c^{PMC}$, and a constant when above $I_c^{PMC}$. The yellow dash-dot line represents $\Delta E_{qt}$+ $E_{cutoff}$ when $I_L < I_c^{NCIF}$, and a constant when above $I_c^{NCIF}$. Three regimes are distinguished by the two critical intensities. (F) Same as (E) obtained from the numerical simulations under the same conditions.

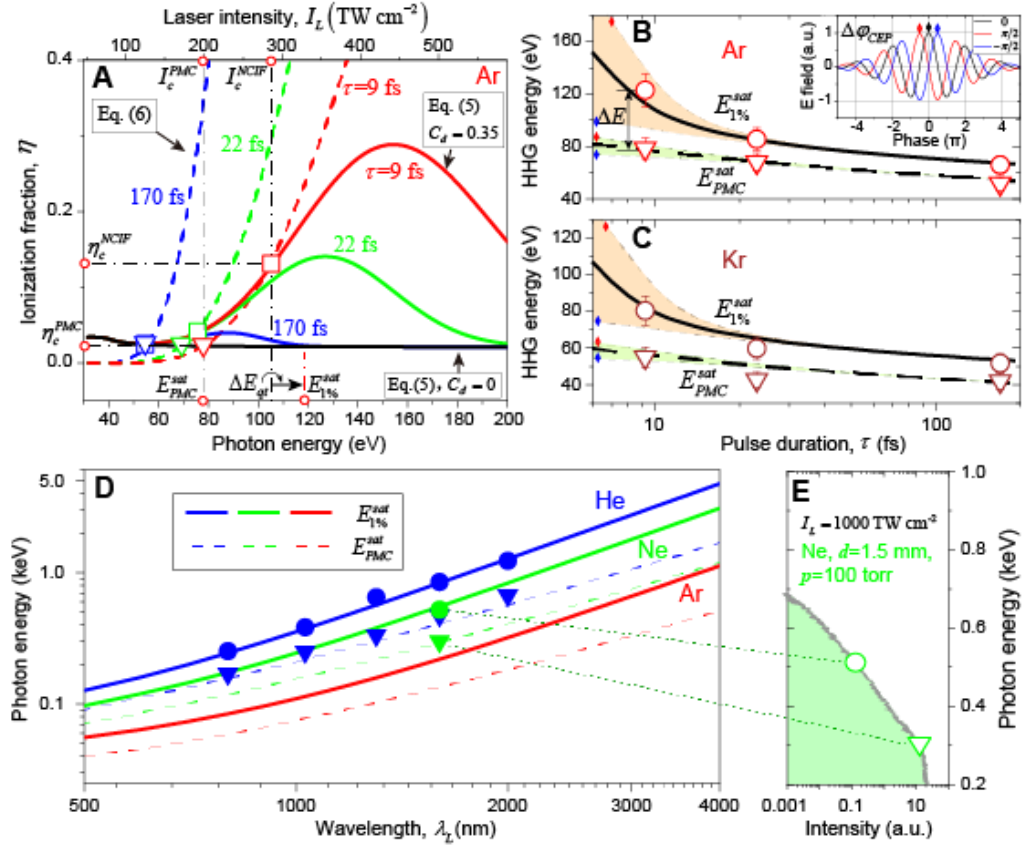

**Figure 3. The NCIF model results.** (A) The results of the NCIF model for HHG in argon under different pulse durations ($\tau$). The solid lines are the results of Eq. (5), with $C_d=0.35$ (red, green and blue) and $C_d=0$ (black). The dashed lines are the results of Eq. (6). The intersections are labeled by open symbols. The open triangles labels the results for PMC and the open squares are for the 1%-intensity energy. (B) The comparison between the experimental results of $E_{PMC}^{sat}$ (open triangles), $E_{1\%}^{sat}$ (open circles) and the NCIF model results (solid and dashed lines) in argon. The shaded area represents the variation of $E_{PMC}^{sat}$ and $E_{1\%}^{sat}$ under different CEP phases ($\Delta\varphi_{CEP}$). Inset: The illustration of the driving field waveforms under different $\Delta\varphi_{CEP}$, with peak-field time labeled. (C) Same as (B) for the results in krypton. (D) The NCIF model results under different wavelengths ($\lambda_L$) in argon, neon and helium. The symbols represent the results obtained from the numerical simulations. (E) A typical numerical spectrum of HHG in neon. The $E_{PMC}$ and $E^{1\%}$ are labeled. The gas pressure is 100 torr, cell length is 1.5 mm and the driving intensity is 1000 TW cm$^{-2}$.

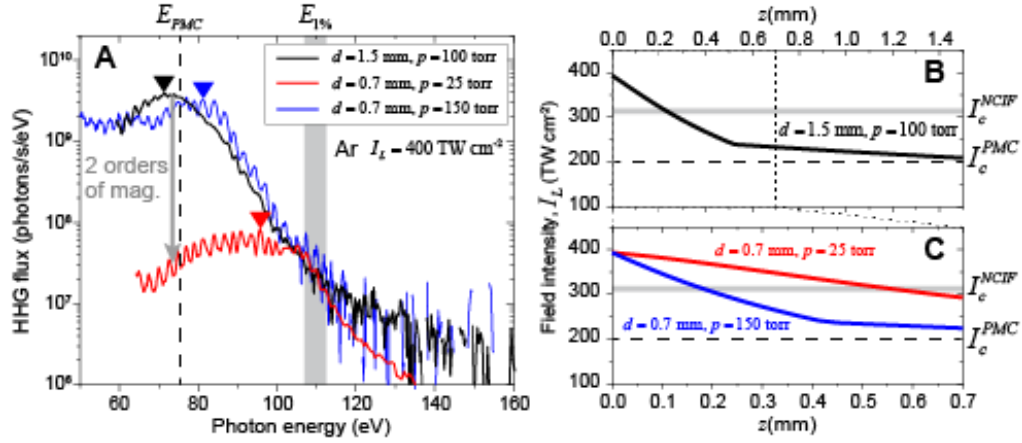

**Figure 4. The nonadiabatic spatial integration effects.** (A) The HHG spectrum in argon obtained under different gas-cell lengths (*d*) and gas pressure (*p*). The driving intensity $I_L$ is fixed to be 400 TW cm$^{-2}$ and the pulse duration $\tau$=9 fs. The energy of $E_{PMC}$ is labeled by solid triangles. (B) The peak intensity of the driving field as a function of the propagation distance in the gas cell (*z*), obtained from the numerical simulation. The critical intensities are labeled for $\tau$=9 fs, $\lambda_L$=1030 nm in argon, obtained from the NCIF model. (C) Same as (B) for different *d* and *p*.